\documentclass[12pt]{article}
\usepackage{amsfonts,amsmath,amsthm,graphicx,mathrsfs}
\usepackage[dvipsnames]{xcolor}
\usepackage[english]{babel}
\usepackage[letterpaper,top=2cm,bottom=3cm,left=3cm,right=3cm,marginparwidth=1.75cm]{geometry}
\usepackage[colorlinks=true, allcolors=blue]{hyperref}

\usepackage[normalem]{ulem}

\title{Arrival Times Versus Detection Times}
\author{
	Sheldon Goldstein\footnote{Departments of Mathematics and Physics, Rutgers University, 110 Frelinghuysen Road, Piscataway, NJ 08854-8019, USA. E-mail: oldstein@math.rutgers.edu},~~
	Roderich Tumulka\footnote{Fachbereich Mathematik, Eberhard-Karls-Universit\"at T\"ubingen, Auf der Morgenstelle 10, 72076 T\"ubingen, Germany. E-mail: roderich.tumulka@uni-tuebingen.de},
	~and Nino Zangh\`\i\footnote{INFN sezione di Genova and Dipartimento di Fisica, Universit\`a di Genova, via Dodecaneso 33, 16146 Genova, Italy. E-mail: zanghi@ge.infn.it}
}

\DeclareMathOperator{\Tr}{Tr}
\newcommand{\CCC}{\mathbb{C}}
\newcommand{\EEE}{\mathbb{E}}

\newcommand{\RRR}{\mathbb{R}}

\newcommand{\tp}{{\mathrm{p}}}
\newcommand{\inter}{{\mathrm{int}}}
\newcommand{\be}{\begin{equation}}
\newcommand{\ee}{\end{equation}}
\newcommand{\va}{{\bf a}}
\newcommand{\vb}{{\bf b}}
\newcommand{\vc}{{\bf c}}
\newcommand{\ve}{{\bf e}}
\newcommand{\vj}{{\bf j}}

\newcommand{\vm}{{\bf m}}
\newcommand{\vn}{{\bf n}}
\newcommand{\vp}{{\bf p}}
\newcommand{\vq}{{\bf x}}
\newcommand{\vu}{{\bf u}}
\newcommand{\vx}{{\bf x}}
\newcommand{\vy}{{\bf y}}
\newcommand{\vz}{{\bf z}}

\newcommand{\vQ}{{\bf X}}

\newcommand{\vnu}{\boldsymbol{n}}
\newcommand{\vsigma}{\boldsymbol{\sigma}}
\newcommand{\vzero}{\boldsymbol{0}}

\newcommand{\sP}{\mathscr{P}}

\newcommand{\qX}{\xi}

\begin{document}
\maketitle

\begin{abstract}
How to compute the probability distribution of a detection time, i.e., of the time which a detector registers as the arrival time of a quantum particle,  is a long-debated problem. In this regard, Bohmian mechanics provides in a straightforward way the distribution of the time at which the particle actually does arrive at a given surface in 3-space in the absence of detectors. However, as we discuss here, since the presence of detectors can change the evolution of the wave function and thus the particle trajectories, it cannot be taken for granted that the arrival time of the Bohmian trajectories in the absence of detectors agrees with the one in the presence of detectors, and even less with the detection time. In particular, we explain why certain distributions that Das and D\"urr \cite{DD19} presented as the distribution of the detection time in a case with spin, based on assuming that all three times mentioned coincide, are actually not what Bohmian mechanics predicts. 

\medskip

\end{abstract}

\section{Introduction}
\label{sec:intro}

In a time-of-arrival experiment, one places detectors along a surface $\Sigma$ in 3-space and asks for the time $T_D$ and place $\vQ_D$ recorded for the detection of a particle. In Bohmian mechanics, a particle has a well defined position  $\vQ(t)$ at any time $t$, so there is a fact even in the absence of detectors about when and where the particle first arrives at $\Sigma$. However, the presence of a detector can change the Bohmian trajectories, even if no detector clicks. After all, as we discuss, the detector may effectively collapse away parts of the wave function that would have influenced the trajectory. That is, possibly $\vQ_{WOD}(t) \neq \vQ_{WID}(t)$, where $WOD$ means \emph{without detector}, $WID$ means \emph{with detector}. As a consequence, possibly
\be\label{TWODneqTWID}
T_{WOD}\neq T_{WID}
\ee
for the first arrival time without detectors, 
\be\label{TWODdef}
T_{WOD}\equiv \inf \{t\geq 0: \vQ_{WOD}(t)\in \Sigma \}
\ee
and likewise with detectors. We write $\sP_{WOD}$ ($\sP_{WID}, \sP_D$) for the probability distribution of $T_{WOD}$ ($T_{WID},T_D$). 

To underscore the importance of these concepts for the paper's comprehension, albeit with some repetition, we will examine them from a different perspective. For ease of visualization, consider the following experiment, which parallels a scenario to be discussed below: within an empty cylinder at some initial time, a particle's wave function $\psi $ is initialized on the bottom base, moving towards the top base. The cylinder's boundaries are impenetrable walls, allowing the quantum dynamics to unfold freely. 
Atop the cylinder sits a detector---a macroscopic device with internal circuitry that triggers upon interaction with the particle, displaying a time $T_D $ on a connected computer screen.

Now, imagine removing the detector from the top base: $T_{WOD} $ represents the time when, according to Bohmian mechanics, the particle first crosses the top base. This time is theoretically computed based on $\psi $, the initial position, and the particle's Hamiltonian. Restore the detector and repeat the experiment with the particle in the same initial state and position. This reintroduces a many-body system involving the detector's constituents in an initial ready state, exactly as in the first scenario. In the first scenario, \( T_D \) represents the numerical result displayed on the computer screen as a consequence of the macroscopic amplification process triggered by the microscopic interaction between the particle and the relevant part of the detector. On the other hand, $T_{WID} $ denotes the time when the particle first crosses the top base, influenced by the many-body wave functions governed by a many-body Hamiltonian, theoretically predicted from the particle's Bohmian trajectory within this many-body system. These three times are clearly conceptually distinct. Our objective in this paper is to explore the relationships among them, if any.

\bigskip 

Das and D\"urr \cite{DD19} computed $\sP_{WOD}$ for a particular setup involving a spin-1/2 particle in an axially symmetric wave guide. Assuming that the presence of the detector would be a ``mild disturbance'' and thus that $T_{WOD}=T_{WID}=T_D$ up to negligible errors, they suggested that the prediction of Bohmian mechanics for $\sP_D$ agrees with $\sP_{WOD}$. 

Recently, we proved in \cite{GTZ23a} that the specific distribution $\sP_{WOD}$ computed in \cite{DD19} cannot be the prediction of Bohmian mechanics for $\sP_D$ because it is not given by a positive-operator-valued measure (POVM), not even approximately. (See Appendix~\ref{app:POVM} for background on POVMs.) Here, we give more detail about the proof and related considerations, including a more detailed analysis of the distributions compatible with a POVM. We provide as well further, alternative arguments for $\sP_{WOD}\neq \sP_D$, including some discussion of the relations among  $T_{WOD}, T_{WID}$, and $T_D$. In particular, we suggest that even $\vQ_{WID}=\vQ_D$ and $T_{WID}=T_D$ can't be taken for granted.

To be sure, there are situations in which the presence of detectors does not significantly influence the trajectories, for example in far-field detection (i.e., as $t\to\infty$, at great distances from the support of the initial wave function) as considered in scattering theory. In that case, 
\be\label{TWODTD}
T_{WOD}=T_{WID}=T_D,
\ee
at least approximately, and $\sP_D$ can be computed by computing $\sP_{WOD}$. Specifically, for $\Sigma$ the sphere around the origin of large radius $R$, the distribution of $T_{WOD}$ is (with small relative error) given by a POVM, in fact a projection-valued-measure (PVM), namely corresponding to the quantum observable $mR/|\hat{\vp}|$ with $\hat{\vp}=(\hat{p}_x,\hat{p}_y,\hat{p}_z)$ the  momentum operator and $m$ the particle's mass.\footnote{Moreover, for large $R$, the joint distribution of $(T_{WOD},\vQ_{WOD})$ is given by the observable $(mR|\hat{\vp}|^{-1},R|\hat{\vp}|^{-1}\hat{\vp})$.}
But in near-field detection, there is no obvious self-adjoint operator that could serve as the observable for $T_D$, which is the origin of the perennial debate about how to compute $\sP_D$. This situation has led in orthodox quantum mechanics to a multitude of guesses about $\sP_D$, as well as to premature claims that it is impossible to determine $\sP_D$ for a detecting surface because of the quantum Zeno effect.

A moral from our discussion is that \emph{detection time is not the same thing as arrival time}. It turns out to be wise to keep in mind the advice often emphasized in orthodox quantum mechanics that we should not take measurement for granted, that the apparatus often plays an active role that must be taken into account. Note, however, that the difference between detection time and arrival time is clearly meaningful and visible in a precise theory such as Bohmian mechanics, whereas in the framework of orthodox quantum mechanics what is happening on the microscopic level is too vague for this to be so.

\bigskip

The remainder of this paper is organized as follows.
In Section~\ref{sec:setup}, we recall the equations of Bohmian mechanics, describe the setup considered by Das and D\"urr in \cite{DD19}, and discuss some reasons for expecting that $\sP_{WOD}\neq \sP_D$.
In Section~\ref{sec:POVMs}, we analyze POVMs on $\CCC^2$, provide alternative proofs of the no-go theorem of \cite{GTZ23a}, and discuss why superluminal signaling would be possible if $\sP_{WOD}=\sP_D$.
In Section~\ref{sec:TWIDTD}, we consider different equivariant equations of motion and discuss their relevance to the question whether $T_D=T_{WOD}$, as well as whether $T_D=T_{WID}$.
In Section~\ref{sec:rem}, we conclude. We also have two appendices. In Appendix~\ref{app:POVM} we review  POVMs and how they arise. In Appendix~\ref{app:B} we respond to a comment of Das and Aristarhov \cite{DA23}.

\section{Arrival Times}
\label{sec:setup}

\subsection{Bohmian Arrival Times}

\paragraph{Bohmian mechanics, a one-parameter family.}
We consider a 1-particle wave function $\Psi=\Psi_t(\vq)$ that is spinor-valued and obeys Schr\"odinger's equation
\begin{equation}\label{Schr}
i\hbar \frac{\partial \Psi}{\partial t} = -\frac{\hbar^2} {2m} \Delta \Psi + V\Psi \,,
\end{equation}
where $V(\vq)$ is a Hermitian $2\times 2$ matrix for every $\vq\in \RRR^3$.
As the Bohmian equation of motion for the particle position $\vQ= \vQ(t)$, two different possibilities are sometimes considered, corresponding to $\lambda=1$ or $\lambda=0$ in the general form
\begin{align}\label{ge}
    \frac{d\vQ}{dt} 
    &= \frac{\hbar}{m} \text{Im} \frac{\Psi^\dagger \nabla \Psi }{\Psi^\dagger\Psi} (\vQ)
    + \lambda\frac{\hbar}{2m} \,\frac{\nabla \times (\Psi^\dagger \vsigma\,\Psi)}{\Psi^\dagger \Psi} (\vQ)\\[3mm]
    &= \frac{\hbar}{m} \text{Im} \frac{\Psi^\dagger\bigl(I_2-i\lambda\vsigma\times \bigr)\nabla\Psi}{\Psi^\dagger\Psi}(\vQ)\,,\label{ge2}
\end{align}
where $\vsigma=(\sigma_x, \sigma_y, \sigma_z) $ is the triple of the Pauli matrices, $\Psi^\dagger \Phi$ denotes the scalar product in the spinor space $\CCC^2$, $\times$ the cross product in 3d, and $I_2$ the $2\times 2$ identity matrix.\footnote{Equation \eqref{ge} has perhaps been considered so far only for $\lambda=0$ or $\lambda=1$. But for the purpose of our analysis we shall allow other real values as well.} In \cite{DD19}, $\lambda=1$ was considered, which arises in the non-relativistic limit of the Bohmian equation of motion associated with the Dirac equation. In fact, $|\Psi|^2$ is equivariant for every $\lambda\in\RRR$. $T_{WOD}$ can be computed for every initial position $\vQ_0$ by solving \eqref{Schr}, \eqref{ge}, and \eqref{TWODdef}. Since $\vQ_0$ is random with distribution $|\Psi_0|^2$, $T_{WOD}$ is random, too, and its distribution can in principle be computed.

Note that even if $V(\vq)$ is real-valued (i.e., a multiple of $I_2$), so that the Schr\"odinger equation \eqref{Schr} does not even couple to spin, then the equation of motion \eqref{ge} still couples to spin for $\lambda\neq 0$, leading to a spin dependence of $T_{WOD}$ and the question whether $T_D$ might depend on the spin as well.

\paragraph{Setup.}
Das and D\"urr \cite{DD19}  considered the real-valued potential $V$ given by
\be\label{Vdef}
V(x,y,z)\equiv \begin{cases}
\infty&\text{if }z<0\\
\tfrac{m}{2}\omega^2(x^2+y^2)&\text{if }z\geq 0 \,,
\end{cases}
\ee
where $\omega>0$ is a constant. This potential does two things: it provides a reflecting wall at $z=0$ and acts as a wave guide that will keep the wave function near the $z$-axis. 

For any unit vector ${\bf n}$ in three-dimensional space, consider the spinor $|{\bf n}\rangle$ in $\mathbb{C}^2$ defined uniquely up to a phase $\theta$ by the condition
$ {\bf n}  = \langle {\bf n} |  \boldsymbol{\sigma} |{\bf n}\rangle
$. Explicitly,
\be\label{chi}
|{\bf n}\rangle = e^{i\theta}\begin{pmatrix}
\cos (\alpha/2)\\
\sin (\alpha/2) \, e^{-i\beta}
\end{pmatrix}
\ee
with $\alpha$ and $\beta$ spherical coordinates of $\vn$, i.e., $\alpha\in[0,\pi]$ the angle between ${\bf n}$ and the $z$-axis and $\beta\in [0,2\pi)$ the angle between $(n_x,n_y)$ and the $x$-axis (azimuthal angle).

The initial wave function is taken to factorize,
\be\label{Psi0}
\Psi_0=|\vn\rangle \otimes \psi_0
\ee
with arbitrary unit spinor $|\vn\rangle$ and a certain fixed choice of $\psi_0$.
The surface considered is
\be\label{Sdef}
\Sigma=\{(x,y,z)\in\RRR^3: z=L\}
\ee
for some  $L>0$. It follows that $\Psi_t=|\vn\rangle\otimes\psi_t$ with $\psi_t$ the solution of the corresponding scalar Schr\"odinger equation, and that $\psi_t(x,y,z)$ vanishes for $z<0$ and all $t$ (but can be non-zero for $z>L$).

\begin{figure}
	\centering
	\includegraphics[width=0.6\textwidth]{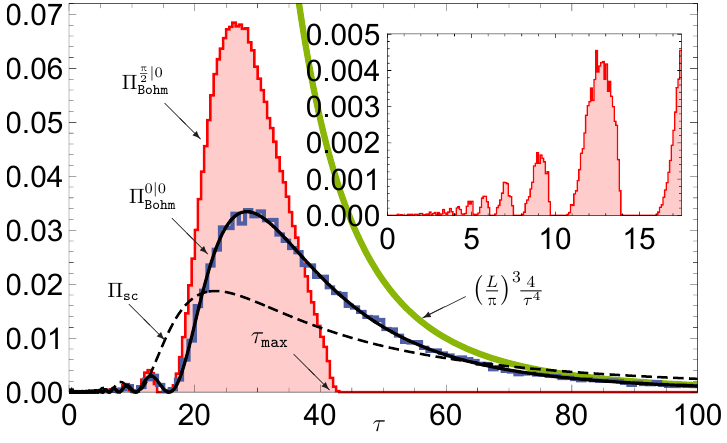}
	\caption{$\sP_{WOD}$ in the setting of \cite{DD19}, here denoted by $\Pi^{\alpha|\beta}_\mathrm{Bohm}$, for two different spinors $|\vn\rangle$; the blue curve ($\alpha=0$) corresponds to $\vn$ parallel to the $z$-axis, the red curve ($\alpha=\tfrac{\pi}{2}$) to $\vn$ perpendicular to the $z$-axis. (Reproduced from \cite{DD19}. The inset shows a detail of the red curve in magnification.)}
	\label{fig:1}
\end{figure}

\begin{figure}
	\centering
	\includegraphics[width=0.6\textwidth]{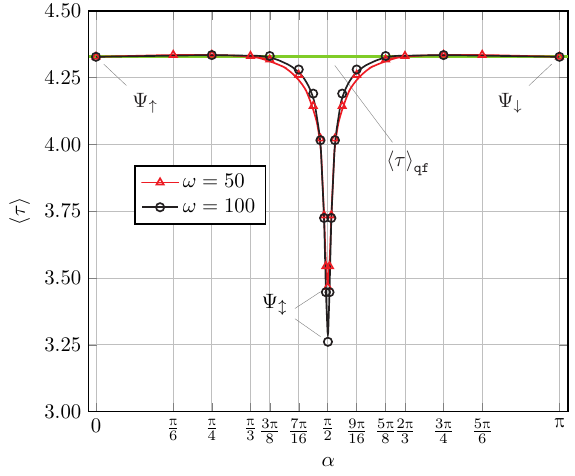}
	\caption{The red and black curves show, for two values of the parameter $\omega$ from \eqref{Vdef}, the expected arrival time $\EEE T_{WOD}$, here denoted by $\langle\tau\rangle$, in the setting of \cite{DD19} as a function of $\alpha$, the angle between $\vn$ and the $z$-axis. (Reproduced from \cite{DD19}. The red and black curves are not related to the red and black curves in Figure~\ref{fig:1}.)}
	\label{fig:2}
\end{figure}

\paragraph{Findings for $\sP_{WOD}$.} Das and D\"urr found that the distribution $\sP_{WOD,\vn}$ of $T_{WOD}$ for given $\vn$, with spin-dependent law of motion \eqref{ge} with $\lambda=1$, depends strongly on the choice of the spinor $|\vn\rangle$. This is visible in Figure~\ref{fig:1} as the difference between the red and the blue curve, and in Figure~\ref{fig:2} as the non-constancy of either the red or the black curve. We will derive below that $\sP_D$ (and its mean) can depend on $\vn$ only in very limited ways that are incompatible with the $\vn$-dependence of $\sP_{WOD}=\sP_{WOD,\vn}$. We also note for comparison that for the spin-\emph{independent} law of motion \eqref{ge} with $\lambda=0$, the trajectories and therefore $T_{WOD}$ are of course independent of $\vn$.

\subsection{The Decoupling Argument}
\label{sec:decouple}

Before we discuss the no-go theorem of \cite{GTZ23a} in Section~\ref{sec:POVMs}, we describe another, much simpler argument to the effect that $T_{WOD}\neq T_D$ in the setting of \cite{DD19}. We will call it the {\it decoupling argument}: 

Assume that the initial wave function $\Psi_0$ of the  particle is of product form \eqref{Psi0} with any unit spinor $|\vn\rangle$ and any normalized scalar function $\psi_0$. One of the remarkable features of $\sP_{WOD}$ is its dependence on $\vn$. Although the Hamiltonian $H_\tp$ of the  particle does not couple to the spin, i.e., is of the form 
\be\label{Htpdecouple}
H_\tp=I_2\otimes \tilde{H}_\tp
\ee
relative to the factorization of Hilbert space into spin and position degrees of freedom, the Bohmian equation of motion \eqref{ge} does couple to the spin, which is why the trajectories (and hence $T_{WOD}$) depend on $|\vn\rangle$. Now we regard the detectors as a quantum system coupled to the  particle by means of an interaction Hamiltonian $H_\inter$. If $H_\inter$ does not couple to the spin of the  particle, i.e., if
\be\label{decouple}
H_\inter=I_2 \otimes \tilde{H}_\inter\,,
\ee
then also the full wave function $\Phi$ of  particle and detectors together at any (late) time will be of the form $|\vn\rangle\:\otimes$ something, where the ``something" does not depend upon ${\bf n}$. Thus the $|\Phi|^2$ distribution of the apparatus's pointer position or records will be independent of $\vn$, so that  $\sP_D$ will as well. This shows that the Bohmian prediction for $\sP_D$ differs from that for $\sP_{WOD}$, so $T_{WOD}\neq T_D$, provided \eqref{decouple} holds. This completes the argument. The mathematical content is that
\be\label{thm1}
\begin{minipage}{.7\textwidth}
{\it assuming \eqref{Htpdecouple} and \eqref{decouple}, $\sP_D$ cannot depend on $\vn$.} 
\end{minipage}
\ee

The decoupling argument shows two things: First, that \eqref{TWODTD} is not an innocent assumption, as it is provably false whenever \eqref{Htpdecouple} and \eqref{decouple} are true. Second, that it is reasonable to expect that the observed distribution in the experiment is independent of $\vn$ when \eqref{Htpdecouple} holds, as \eqref{decouple} sounds like a reasonable assumption, one that would be satisfied, at least to a high degree of approximation, for many apparently reasonable detection procedures. However, since observations in actual experiments include relativistic effects in principle, and since the relativistic Hamiltonian of an electron (i.e., the Dirac Hamiltonian) does not decouple from spin as in \eqref{Htpdecouple} (and not either as $I_4 \otimes \tilde{H}_\tp$ for $\CCC^4$ as the spin space), the relativistic (and thus also the observed) distribution for $T_D$ should not (exactly) be independent of $\vn$. On the other hand, the no-go theorem of \cite{GTZ23a} does not assume \eqref{Htpdecouple} or \eqref{decouple}, which is why it provides a relevant piece of information that goes beyond the decoupling argument.

\subsection{How the Presence of Detectors Changes Trajectories}
\label{sec:collapse}

Let us flesh out our statement that the presence of a detector can change the Bohmian trajectories, even if it does not click. 

Consider a double-slit experiment with a detector in the left slit. It will effectively collapse the 1-particle wave function to a wave packet that passed either through the left or the right slit, depending on whether the detector clicked or not. A wave that passed through one slit only will lead to different Bohmian trajectories (no interference fringes, possibility to cross the symmetry plane between the slits) than a wave that passed through both. In this example, the presence of the detector perhaps has only a minor effect on the arrival \emph{time} at the screen, but a major effect on the arrival \emph{place} at the screen. But there are three aspects to note: first, that the effect on the trajectories is not a mild disturbance, but a drastic change of the trajectory; second, that using better detectors will not make the change less drastic; and third, once we realize that the presence of detectors can drastically change the trajectories, we have to expect that in another case also the arrival \emph{time}  would be affected. 

Since in the double-slit experiment with a detector in a slit, the Bohmian trajectory is changed \emph{beyond} the plane containing the detector and the two slits, one might wonder whether, in general, the trajectory is not changed \emph{before} the surface $\Sigma$ containing detectors. This cannot be taken for granted: even neglecting that a part of the wave might be reflected from a detecting surface, we notice that in the free evolution of $\psi$ in the absence of detectors on $\Sigma$, parts of $\psi$ beyond $\Sigma$ can in general also propagate back to the region before $\Sigma$ and contribute there. Thus, if parts of $\psi$ get collapsed away when passing $\Sigma$, $\psi_t$ in the region before $\Sigma$ will be different from what it would have been in the free evolution. As a consequence, also trajectories will be different from what they would have been in the free evolution.

\section{POVMs}
\label{sec:POVMs}

In this section, we discuss several ways of proving the no-go theorem mentioned above, which asserts \cite{GTZ23a} that \emph{there is no possible experiment {(for which the experimental design does not depend upon $\vn$)} that would, when applied to a particle in the setup of \cite{DD19} with $\Psi_0=|\vn\rangle \otimes \psi_0$, have outcome with distribution $\sP_{WOD,\vn}$ for every $\vn$}.  
The theorem means, in other words, that there can be no experiment or procedure that yields the arrival time $T_{WOD}$ unless the procedure itself directly incorporates information about the direction $\vn$ of the spin state; 
it shows in particular that if the experiment consists of putting on $\Sigma$ the kind of device commonly regarded as a detector, however it might work, the distribution $\sP_D$ of its outcome $T_D$ will not in general agree with $\sP_{WOD}$.

By experiment we shall henceforth always mean one  whose design is independent of $\vn$ (respectively $\Psi_0$). By way of contrast, suppose we were to measure the position of the particle at the initial time and then use our knowledge of the quantum state of the particle to compute (what would have been) its trajectory and thereby its arrival time $T_{WOD}$. Such a procedure would not count as an experiment according to this stipulation. 

We start from the known fact that for every experiment that can be carried out on a system with arbitrary initial wave function $\Psi_0$, there is a POVM $E_0(\cdot)$ such that the outcome statistics are given by $\langle\Psi_0|E_0(\cdot)|\Psi_0\rangle$. This fact is a theorem in Bohmian mechanics \cite{DGZ04}, \cite[Sec.~5.1]{Tum22}. In particular, $\sP_D$ must be given by some POVM $E_0(dt)$. But, as  we will show, $\sP_{WOD}$ is not given by a POVM.

\subsection{Spin POVM}

Since $\Psi_0$ factorizes according to \eqref{Psi0} into a spinor $|\vn\rangle$ and a scalar function $\psi_0$, if we regard $\psi_0$ as fixed and $|{\bf n}\rangle$ as variable, then the spin dependence of the outcome statistics is captured by the  POVM 
\be\label{spinPOVM}
E(dt)=\langle \psi_0|E_0(dt)|\psi_0\rangle, 
\ee
where the inner product is a partial inner product taken in the position degrees of freedom but not those for the spin. Thus $E(dt)$ is  a \emph{spin POVM}, a POVM on $\RRR$ acting on $\CCC^2$; as such it is given by $2\times 2$ matrices.  It follows that for $\Psi_0=|\vn\rangle\otimes\psi_0$, the distribution of the detection time is 
\be
P_\vn(dt)\equiv \langle \vn |E(dt)|\vn\rangle\,.
\ee
For example, the most familiar spin POVM is for the measurement of the $z$-component $\sigma_z$ of spin for a spin-1/2 particle, which is given by 
\be
E=\tfrac12\Bigl(\delta(t-1)+\delta(t+1)\Bigr)I_2+\tfrac12\Bigl(\delta(t-1)-\delta(t+1)\Bigr)\sigma_z
\ee
with $I_2$  the $2\times2$ identity.

In the following, we point out several properties of distributions arising from spin POVMs that are incompatible with the $\vn$-dependence of $\sP_{WOD}$ found in \cite{DD19}; each of them proves the no-go theorem asserting that $\sP_{WOD}\neq \sP_D$. Such a property was first pointed out in \cite{GTZ23a}.

\subsection{Properties of Distributions from POVMs}

\paragraph{Trace property.} Since $|\vn\rangle$ and $\left|-\vn\right\rangle$ form an orthonormal basis of spin space $\CCC^2$, and since the trace of an operator is the same in any basis, we have that for any  spin POVM $E(\cdot)$,
\be\label{tracegeneral}
\left\langle -\vn\right|E(\cdot)\left|-\vn\right\rangle + \langle \vn|E(\cdot)|\vn\rangle = \Tr E(\cdot)
~~\text{\it is independent of $\vn$.}
\ee
In particular,
\be\label{tracethm}
P_{\vn}(dt)+P_{-\vn}(dt)
~~\text{\it is independent of $\vn$.}
\ee 
Now let $\vx,\vy,\vz$ be the unit vectors in the $x,y,z$ direction. As a consequence of the axial symmetry of the setup of Das and D\"urr,  $\sP_{WOD,\vn}$ is independent of $\beta$; thus
\be\label{sPx}
\sP_{WOD,-\vx}=\sP_{WOD,\vx}\,.
\ee
They also proved that
\be\label{sPz}
\sP_{WOD,-\vz}=\sP_{WOD,\vz}\,.
\ee
If $\sP_{WOD}$ of \cite{DD19} came from a POVM $E$, then we would have that $\sP_{WOD,\vn}=P_\vn$ for all $\vn$, so that $P_{-\vx}=P_{\vx}$ and $P_{-\vz}=P_{\vz}$. But then, by \eqref{tracethm}, 
\be
P_{\vx}=\tfrac{1}{2}(P_{\vx}+P_{-\vx})=\tfrac{1}{2}(P_{\vz}+P_{-\vz})=P_{\vz}
\ee
and hence $\sP_{WOD,\vx}=\sP_{WOD,\vz}$, in contradiction with the strong difference between them found in \cite{DD19} (clearly visible as the difference between the red and blue curves in Figure~\ref{fig:1}). This is our first proof of the no-go theorem.

In fact, there cannot even exist a spin POVM $E$ that agrees with $\sP_{WOD,\vn}$ for merely the three directions $\vn=\vx, -\vx$ and $\vz$. This is so because in this case the trace property would imply that $P_\vz\leq 2P_\vx$, which is clearly incompatible with what we see in Fig. \ref{fig:1}. This is our second proof of the no-go theorem.

\paragraph{Remark.} Apart from arrival time questions, Eq.~\eqref{tracegeneral} can be used to show that the spin state $|\vn\rangle$ itself (or the associated direction $\vn$) is not measurable; that is, there is no experiment that could, upon the input of a particle with arbitrary spin state $|\vn\rangle$, yield $\vn$ as the result. Indeed, if such an experiment existed, it would be associated with a POVM $F(d\vu)$ on the unit sphere $\{\vu\in\RRR^3:|\vu|=1\}$ acting on $\CCC^2$ such that $\langle \vn|F(d\vu)|\vn\rangle=\delta(\vu-\vn)\, d\vu$. 
But $\delta(\vu-\vn)+\delta(\vu+\vn)$ clearly depends upon $\vn$, so that by \eqref{tracegeneral} no such POVM exists.

\paragraph{Relation to Pauli matrices.} 
Any Hermitian $2\times 2$ matrix $M$ can be written in the form
\be\label{Msigma}
M=m_0 I_2 + \vm \cdot \vsigma 
\ee
with some real $m_0$ and some real 3-vector $\vm$. We thus have that
\be\label{Esigma}
E(dt) = e_0(dt) I_2 + \ve(dt) \cdot \vsigma
\ee
and thus
\be\label{Pe}
P_\vn(dt)=e_0(dt)+\ve(dt)\cdot \vn\,
\ee
for suitable $e_0$ and $\ve=(e_x,e_y,e_z)$. 

Note that by \eqref{Pe} $e_0(dt)=\frac12 (P_{\vn}(dt)+P_{-\vn}(dt))$ (yielding \eqref{tracethm} once again). Thus $e_0$ must be a probability measure. As a consequence, the signed measures $e_\nu$ (for $\nu=x,y,z$) must have vanishing total mass, $\int_{-\infty}^\infty e_\nu(dt)=0.$ Moreover,  any such $e_0$ and $\ve$ correspond via \eqref{Esigma} to a spin POVM if and only if $\ve^2\leq e_0^2$, since, as it is not hard to see,  this amounts to positivity.

\paragraph{Axial symmetry.}
Now take the axial symmetry of the setup into account. Then  were $\sP_{WOD}$ given by a POVM $E$, $E$ itself would have to be axially symmetric, so that its dependence on $\vsigma$ would be  via only $\sigma_z$. Thus it would be of the form
\be\label{Eaxial}
E(dt) = e_0(dt)I_2 + e_z(dt) \sigma_z \,,
\ee
which leads to 
\be\label{ap}
P_{\vn}(dt)=e_0(dt) +  e_z(dt)\cos{\alpha}.
\ee
Setting $\alpha=\pi/2$ we would thus have that $e_0 = P_{\vx}$;  setting $\alpha=0$ and $\alpha=\pi$ and adding we would obtain that $2e_0=P_{\vz} + P_{-\vz}$.  But this---or \eqref{tracethm}---implies that $P_{\vz} + P_{-\vz}= 2P_{\vx}$, and hence that 
$\sP_{WOD,\vz} + \sP_{WOD,-\vz} =2\sP_{WOD,\vx}$, in contradiction with what we see in Figure \ref{fig:1}. This is our third proof of the no-go theorem. (Note that the positivity condition for an axially symmetric POVM \eqref{ap} is that $|e_z|\leq e_0$.)

\paragraph{Condition for spin independence.} It follows  from \eqref{Pe}  that for any unit vector $\va$, $P_{\va}-P_{-\va}=2\ve\cdot\va$, so the vector $\ve=\ve(dt)$ can be expressed for any orthonormal basis $\va,\vb,\vc$ as
\be\label{eDeltaP}
\ve = \tfrac{1}{2}(P_{\va}-P_{-\va})\va +\tfrac{1}{2}(P_{\vb}-P_{-\vb})\vb +\tfrac{1}{2}(P_{\vc}-P_{-\vc})\vc\,.
\ee
Therefore,
\be\label{thm4}
\begin{minipage}{.7\textwidth}
{\it if $P_{-\va}=P_{\va}$, $P_{-\vb}=P_{\vb}$, and $P_{-\vc}=P_{\vc}$ for orthonormal vectors $\va,\vb,\vc$, then $\ve(dt)=\vzero$, and $P_\vn$ is independent of $\vn$.}
\end{minipage}
\ee
Now use \eqref{sPx} and the fact that,  by axial symmetry, $\sP_{WOD,-\vy}=\sP_{WOD,\vy}$, as well as \eqref{sPz}; if $\sP_{WOD}$ came from a POVM, then $\sP_{WOD,\vn}=P_\vn$ for all $\vn$, so $P_{-\vx}=P_{\vx}$, $P_{-\vy}=P_{\vy}$, and $P_{-\vz}=P_{\vz}$. By \eqref{thm4}, $P_\vn$ must then be independent of $\vn$, so $\sP_{WOD,\vn}$ is then independent of $\vn$, but we know it is not. 
This is our fourth proof of the no-go theorem.

\paragraph{Sinusoidal dependence.} Let us go back to \eqref{Pe}. In terms of the spherical coordinates $\alpha,\beta$ mentioned in Section~\ref{sec:setup}, \eqref{Pe} becomes a combination of the first (and 0-th) spherical harmonics:
\be\label{Pnsin0}
P_{\vn}(dt) = e_0(dt)+ e_x(dt)\sin\alpha\cos\beta+e_y(dt)\sin\alpha\sin\beta + e_z(dt)\cos\alpha.
\ee
If this is to agree with $\sP_{WOD,\vn}$, which is axially symmetric, it must  be independent of $\beta$ and thus of the form
\be\label{Pnsin1}
P_{\vn}(dt) = e_0(dt)+ e_z(dt)\cos\alpha,
\ee
as we argued earlier.\footnote{A perhaps slightly more convenient formula for $P_{\vn}$ in the case of axial symmetry is obtained by expanding the POVM $E$ \eqref{Eaxial} in terms of $I_2-\sigma_z$ and $\sigma_z$ instead of $I_2$ and $\sigma_z$:
\be\label{E}
E(dt) = A(dt)(I_2-\sigma_z)+ B(dt)\sigma_z,
\ee
so that
\be\label{Pnsin2}
P_{\vn}(dt) = A(dt)(1-\cos\alpha)+B(dt)\cos\alpha.
\ee
Setting $\alpha=0$ and $\pi/2$ we see that $A=P_{\vx}$ and $B=P_{\vz}$, and \eqref{Pnsin2} becomes
\be\label{Pnsin3}
P_{\vn}(dt) = P_{\vx}(dt)(1-\cos\alpha)+P_{\vz}(dt)\cos\alpha.
\ee}
But the dependence of $\sP_{WOD,\vn}(dt)$ on $\alpha$ is not of the form \eqref{Pnsin1}. This can perhaps be seen most easily by considering the expected detection time.

\paragraph{Expected detection time.}  It follows from \eqref{Pnsin1} that were $\sP_{WOD,\vn}$ given by a POVM,  also the expectation $\EEE T_D$ of $T_D$  would depend on $\alpha$ in a sinusoidal way:  We would have that
\begin{align}
\EEE T_D
&= \int_0^\infty t \, P_\vn(dt)\\
&= \int_0^\infty t \, e_0(dt)+\cos\alpha\int_0^\infty t \, e_z(dt)\\[2mm]
&= \tau_0  + \tau_z\cos\alpha\label{et}
\end{align}
with $\tau_\nu\equiv \int_0^\infty t \, e_\nu(dt)$ (for $\nu=0,z$).
Compare this to the graph of the dependence of $\EEE T_{WOD}$ on $\alpha$ found in \cite{DD19}, reproduced here as Figure~\ref{fig:2}.  They are  clearly incompatible. This is our  fifth proof of the no-go theorem.

\bigskip

The sinusoidal dependence of $\EEE T_D$ on $\alpha$ can also be obtained from \eqref{Eaxial}:
\be
\EEE T_D=\langle \vn |M|\vn \rangle\,,
\ee
where the matrix $M$ is the mean of the POVM $E(dt)$,
\be
M=\int_0^\infty t \, E(dt) =\tau_0I_2+ \tau_z\sigma_z\,, 
\ee
yielding \eqref{et}.

\paragraph{Approximate measurement.}   We've seen that there is no POVM that will make $P_\vn$ coincide with $\sP_{WOD,\vn}$ even for at least the 3 directions $\vz,\vx,-\vx$ and certainly not for all 4 directions $\vz,-\vz,\vx,-\vx$. This remains true when we replace ``coincide'' with ``coincide approximately'': 
\begin{equation}\label{4approx}
\begin{minipage}{.7\textwidth}
{\it For every spin POVM $E$ there is a direction $\vn$ among $\vz,-\vz,\vx,-\vx$ such that} 
$$\|\sP_{WOD,\vn}-P_\vn\| \geq \|\sP_{WOD,\vx}-\sP_{WOD,\vz}\|/2\,.$$
\end{minipage}
\end{equation}
Since the right-hand side as computed in \cite{DD19} is a non-negligible quantity, this relation means that $\sP_{WOD,\vn}$ cannot be close to $P_\vn$ for all 4 directions. 

To verify \eqref{4approx}, 
consider the triangle inequality in the version
\be
\|D_\vz + D_{-\vz} - D_\vx - D_{-\vx}\| \leq \|D_\vz\|+ \|D_{-\vz}\| + \|D_\vx\|+ \|D_{-\vx}\|
\ee
for $D_\vn = \sP_{WOD,\vn}-P_\vn$. By \eqref{tracethm}, the contributions from $P$ cancel on the left-hand side. By \eqref{sPx} and \eqref{sPz}, the left-hand side equals $2\|\sP_{WOD,\vx}-\sP_{WOD,\vz}\|$. Finally, it can't be the case that  each of the 4 summands on the right-hand side  is   less than a quarter of the left-hand side, which completes the proof of \eqref{4approx}. (A similar  statement, involving the positive part of $\sP_{WOD,\vz}-2\sP_{WOD,\vx}$, could be made for just the 3 directions.)

\subsection{Previous Results}
 
To put the no-go theorem into context, we mention a couple of similar previous results. 

First, in \cite{DGZ04}, it was shown that, according to Bohmian mechanics, there is no experiment that could measure the velocity $d\vQ/dt$ of a given single particle.\footnote{Measuring its position and then using the guiding equation to compute the velocity requires knowledge of the quantum state and hence does not count as an experiment according to our stipulation.}  One method of proof is to show that the distribution of $d\vQ/dt$ is not given by a POVM; the situation is different with the {\it asymptotic} (long-time average) velocity, which is measurable. 

Second, Vona, Hinrichs, and D\"urr \cite{VHD13} already studied whether $\sP_{WOD}$ is given by a POVM. Their considerations are not limited to the setting of \cite{DD19}, and many of them can be applied to rather arbitrary $V$, $\Sigma$, and $\Psi_0$ (evolving in $\RRR^d$, with or without spin). They considered the set $S$ of initial wave functions $\Psi_0$ obeying
the so-called current positivity condition
\be\label{CPC}
\vnu(\vq)\cdot \vj(t,\vq)
\geq 0~~~\forall\vq\in \Sigma ~ \forall t\geq 0\,,
\ee
where $\vnu(\vq)$ is the outward unit normal vector on $\Sigma$ at $\vq$ and $\vj$ is given by $\Psi^\dagger \Psi$ times the right-hand side of \eqref{ge} or \eqref{ge2}. (They considered $\lambda=0$ but any fixed $\lambda$ could be considered instead.)
From this condition it follows that the distribution of $(T_{WOD},\vQ_{WOD})$ has density $\vnu\cdot\vj$ and is thus given by an operator-valued measure. However, the operators need not be positive, since linear combinations of wave functions in $S$ may fail to be in $S$. 
Of course, an \emph{operator-valued measure} is not sufficient, as the outcome of an experiment always has distribution given by a \emph{positive-operator-valued measure}. Vona et al. established the following  result: Let ($V$ and) $\Sigma$ be given, and let $S_0$ be a set of normalized wave functions satisfying \eqref{CPC}.  Suppose there are $\Psi,\Phi\in S_0$ such that, with $\Psi_{\pm}:=(\Psi\pm\Phi)/\|\Psi\pm\Phi\|$, we have that $\Psi_+\in S_0$ while $\Psi_-$ satisfies
\be
\int_\Sigma d^2\vq~ \vnu(\vq) \cdot \vj_{\Psi_-}(t_0,\vq)<0
\ee
for some $t_0\geq 0$. Then there is no POVM $E_0$ such that $\langle \Psi_0|E_0(\cdot)|\Psi_0\rangle$ agrees with $\sP_{WOD,\Psi_0}$ 
for all $\Psi_0\in S_0$. Moreover there are, at least for $V=0$ and $\Sigma$ a plane,  $\Psi$ and $\Phi$  in $S$ as just described, thus establishing  that there is at least one $\Psi_0\in S$ for which $T_{WOD}\neq T_D$.

Our no-go result and that of Vona et al.\ both concern the non-existence of a POVM for $\sP_{WOD}$, but for different classes of wave functions. Vona et al.'s is for wave functions obeying the current positivity condition  \eqref{CPC} while ours concerns wave functions of the form  $|\vn\rangle \otimes \psi_0$ with any $\vn$, most of which don't  obey \eqref{CPC}. The two results thus complement each other.

\subsection{POVMs and Measurability}

When investigating whether in Bohmian mechanics a quantity $\qX$ (such as the first arrival time at $\Sigma$) is measurable, i.e., whether there is an experiment whose outcome is $\qX$---an experiment that measures $\qX$---we have used that it is a necessary condition that the probability distribution of $\qX$ be given by a POVM. Is this condition also sufficient? No. Here is why. Suppose there is a POVM $E$ such that we have, for the probability distribution of $\qX$ when the system is in state $\Psi$, that $\mathrm{Prob}_\Psi(\qX\in B)=\langle\Psi|E(B)|\Psi\rangle$ for all wave functions $\Psi$ and all sets $B$. There  then remain two possible obstacles to the existence of an  experiment that measures $\qX$: 
\begin{enumerate}
\item Maybe no appropriate apparatus leading to $E$ can be built. To be sure, any POVM $E$ on a system Hilbert space can be expressed in the form
\be\label{POVMformula}
E(B)=\langle \Phi_0  |U^\dagger P(F^{-1}(B)) U|  \Phi_0\rangle\,,
\ee
where the inner product is a partial inner product in suitable apparatus variables, $\Phi_0$ is a suitable ready state of the apparatus, $U$ a unitary operator on the Hilbert space of system and apparatus together, $P$ the position PVM, and $F$ the function that yields the outcome from the final configuration. Nonetheless, it may not be possible to physically realize such $U, \Phi_0$, and $F$. 
\item The other obstacle is that, even if the POVM of some experiment equals $E$, the individual outcome $Z$ of the experiment need not agree with $\qX$. After all, it is only guaranteed that the random variables $\qX$ and $Z$ have the same distribution, but that does not entail they are equal. 

For example, consider a simple harmonic oscillator in $d\ge 2$ space dimensions, with Hamiltonian $H$. Then, since all eigenvalues of $H$ are integer multiples of the ground state energy $\varepsilon_0$ for $d=1$, the time evolution operator $\exp(-iHt/\hbar)$ yields the identity for $t=t_0:=2\pi\hbar/\varepsilon_0$. Thus we have that the wave function $\Psi_{t_0}$ of the system at time $t_0$ is the same as the wave function $\Psi_0$ at time 0, and hence similarly for the distributions of the  (Bohmian)  positions $\vQ(t_0)$ and $\vQ_0$ at those times. Thus the distribution of $\qX=\vQ(t_0)$ is given by a POVM, the position POVM (given by the spectral projections for the position operator). Moreover, a measurement of  $\vQ_0$ is an experiment the distribution of whose result $Z\ (=\vQ_0)$ is given by this POVM. But, nonetheless, this experiment   measures, not  $\vQ(t_0)$, but  $\vQ_0$ instead. Indeed, if $d>1$, the Bohmian position for the harmonic oscillator at time $t_0$  is in general different from the Bohmian position at time $0$, even though the wave function at these times is the same.
 
On top of that, here  is a trivial example, involving a POVM acting on a one-dimensional subspace: Regardless of whether or not  $\qX$ is   measurable,  and whatever the distribution $\sP_{\qX}$ of $\qX$ for $\Psi_0$, there trivially exists a POVM $E$, namely $E=\sP_{\qX} I$, on the 1d subspace $\CCC\Psi_0$, that reproduces $\sP_{\qX}$; and there trivially exists an experiment with this POVM, viz., running a random number generator that creates a number $Z$ with distribution $\sP_{\qX}$. Since $Z$ is generated independently of $\qX$, the two will in general not be equal.
\end{enumerate}

\subsection{Superluminal Signaling}
\label{sec:superluminal}

As has been pointed out by Das and Maudlin \cite{DM22}, it is an easy consequence of the spin dependence of $\sP_{WOD}$ in the setup of \cite{DD19} that the hypothesis $\sP_D=\sP_{WOD}$ would imply the possibility of superluminal signaling. Here is how. 

\paragraph{Procedure.} For sending a bit $X\in\{0,1\}$ from Alice to Bob at spacelike separation, a pair of particles is prepared (as for a Bell experiment) in the singlet state $\propto\bigl| \uparrow \downarrow \bigr\rangle - \bigl| \downarrow \uparrow \bigr\rangle$, then one particle is transported to Alice's lab and one to Bob's. Depending on whether $X=0$ or $X=1$, Alice carries out a Stern-Gerlach experiment on her particle in either the $z$- or the $x$-direction.  Because of the rotational invariance of and the perfect anti-correlations in the singlet state, this will result in a collapsed wave function of Bob's particle that is $|\pm\vz\rangle$ (where each sign has probability $1/2$) if $X=0$ or $|\pm\vx\rangle$ if $X=1$. Now Bob carries out an arrival-time experiment in the setup of \cite{DD19} on his particle. As mentioned  (see \eqref{sPx} and \eqref{sPz}), $\sP_{WOD}$ does not depend on the random sign;  it does, however,  depend on whether the spin direction is $\vz$ or $\vx$, as shown by the blue and red curves in Figure~\ref{fig:1}. So, if $\sP_D$ were equal to $\sP_{WOD}$, then Bob would obtain probabilistic information about $X$; if Alice and Bob repeat the procedure 100 times with the same value of $X$, Bob could decide correctly with high probability whether his observations are distributed according to the blue or the red curve, and thus determine the message $X$ that Alice wanted to send.

When one realizes that a hypothesis implies the possibility of superluminal signaling, one should become very skeptical of the hypothesis, not just because nobody has been able to experimentally achieve it, but also because there is a general proof that it is impossible in Bohmian mechanics \cite{Tau67,Bell76,Eb,GC}, \cite[Sec.s 5.5.9 and 7.6.2]{Tum22}. On top of that, we can analyze the procedure in terms of POVMs; this is the content of the next paragraph.

\paragraph{Does It Work?} We can easily specify the necessary and sufficient condition for the  impossibility of superluminal signaling by means of an experiment of the sort just described, one allowing for the choice of any pair $\vn_1$ and $\vn_2$ of directions and not just $\vz$ and $\vx$: It is that
\be\label{condsignal}
\sP_{D,\vn_1}+\sP_{D,-\vn_1} = \sP_{D,\vn_2}+\sP_{D,-\vn_2}
\ee
for any pair of unit vectors $\vn_1$ and $\vn_2$. After all, $\tfrac{1}{2}(\sP_{D,\vn}+\sP_{D,-\vn})$ is the detection time distribution that Bob will observe if Alice uses the direction $\vn$. As we have noted already in \eqref{tracethm}, \eqref{condsignal} must hold because $\sP_D$ must be given by a spin POVM.

\section{$T_{WID}$ versus $T_D$}
\label{sec:TWIDTD}

It may sound innocent to assume that a detector should click when and where the particle actually arrives, which would imply that
\be\label{TDTWID}
T_D=T_{WID}\,.
\ee
And in classical mechanics, this is the case because the interaction between particle and detector is local. But in Bohmian mechanics, the structure is more involved because, although the interaction term in the Hamiltonian is still local, it directly affects only the (entangled) wave function, which in turn affects the motion of the particles; so it is not obvious whether \eqref{TDTWID} is correct. More precisely, it is not clear whether there exists a device $D$---a potential detector---for which \eqref{TDTWID} holds. It might well be that for any potential detector $D$, $T_{WID}$ is like the spin state $|\vn\rangle$, or the wave function, in that it is measurable, neither by $D$ itself, so that \eqref{TDTWID} 
is violated, nor by any other device. After all, $T_{WID}$ represents the Bohmian arrival time of a particle interacting with a certain environment (the detector), while $T_D$ is the outcome, registered by the device, produced by the interaction, which is conceptually and physically quite different.  Therefore, the  comparison  is similar to the earlier comparison between $T_{WOD}$ and $T_{D}$, with the distinction between the comparisons being the presence or absence of a certain environment.  
(Nonetheless, it could be the case that there is a fundamental difference between measurement of $T_{WID}$ and $T_{WOD}$. The latter can in general not be measured even approximately, whereas for the former, though there may be limitations, it might well be that approximate measurement is always possible; see also \cite{DBD20,Tum24}.)

\subsection{Comparison to Position Measurements}

It should be observed that this state of affairs is not confined to time measurements alone; even when it comes to position measurements, a discrepancy may arise between the actual position in the presence of a detector and the position recorded by the detector. As emphasized in the introduction, even the assumption that $\vQ_{WID}=\vQ_D$ should not be accepted without scrutiny.  In Bohmian mechanics there is no reason for exempting position or time from adhering to the principle that quantum measurement outcomes are the joint result of the combined behaviors of both the system and the apparatus, a principle which is indeed a direct consequence of the  Bohmian analysis of quantum measurements \cite{DGZ04}.

We should also note that for most potential measurements of a quantity $\qX,$ there is no difference between $\qX_{WOD}\ (=\qX)$ and $\qX_{WID}$, so that the measurability of $\qX$ boils down to whether $\qX_{WID}=\qX_D$ (for some device $D$). This will be so whenever the time evolution plays no role in the definition of $\qX$,  unlike for the case of arrival times. For example, this is the case when $\qX$ is the position of a particle, in quantum state $\Psi$, at a given time $t$, say, time $t=0$. The presence or absence of a detector is not relevant to the meaning of $\qX$, though it is of course relevant to the position of the particle at later times.

So let's consider the simple case when $\qX$ is the position of a particle at a given time. It is generally believed, with good reason, that this can be measured with unlimited precision. But even in this case,  one could take a step further and argue that the measurability of position is a contingent feature of the non-relativistic framework within which the theory is formulated. Consider, for instance, a relativistic scenario, specifically involving the measurement of the position of a Dirac electron. Suppose, as is widely believed, that in the one-particle Hilbert space of the electron, comprising exclusively positive energy states, there is an inherent limitation on how narrow the wave function can be.
This restriction is intuitively understandable since a delta function cannot be exclusively formed with positive energy wave functions. 
Then the minimal width of the wave function (and thus, by Born's rule, the minimal possible inaccuracy of an experimenter's knowledge of the electron's position) would occur on a scale approximately corresponding to its Compton wavelength. 
Consequently, any putative detector probing below that scale would register a specific position that typically deviates systematically from the actual position of the electron. 

\subsection{Different Equations of Motion}
\label{sec:motion}

It has long been known that different equations of motion, regarded here as different possible laws of nature, can all leave the Born distribution $|\Psi|^2$ equivariant while leading to different trajectories.  Any such   law of motion, as part of a universal theory, leads to the same probability distribution as in Bohmian mechanics and orthodox quantum mechanics for any experimental outcome. That is because we  should   apply this law of motion to the joint configuration of both the observed system and   the apparatus  to   conclude that the joint configuration of the apparatus and the  system at any time $\tau$ after the experiment has begun, say at time 0, will be $|\Psi_\tau|^2$ distributed; since it can be arranged that the outcome can be read off from the apparatus configuration for some $\tau>0$, the distribution of the outcome must agree with the one obtained from Bohmian mechanics or orthodox quantum mechanics. That is, the theory with this law of motion is empirically equivalent to Bohmian mechanics and orthodox quantum mechanics (even though it need not lead to the same individual outcome as Bohmian mechanics when starting from the same initial configuration). 

Of course, this consideration includes the distribution $\sP_D$ of the detection time of a particle in an arrival-time experiment.
But when applied to the particle alone, a different law of motion leads to generically different arrival times $T_{WOD}$ on the surface $\Sigma$, and in fact to different distributions $\sP_{WOD}$. Thus, for most of such laws of motion, $\sP_D\neq \sP_{WOD}$.

\subsubsection{Examples}

Here are four different examples of such laws of motion:

\begin{itemize}
\item For the Bohmian equation of motion for the position $\vQ(t)$ of a non-relativistic spin-$\tfrac{1}{2}$ particle, there are the possibilities indexed by $\lambda\in\RRR$ in \eqref{ge}. As mentioned already in Section~\ref{sec:setup}, $\sP_{WOD}$ is independent of $\vn$ for $\lambda=0$ but not for $\lambda\neq 0$.

\item Nelson's stochastic mechanics \cite{Nel85,Gol87} posits a stochastic particle motion given by a diffusion process. The process belongs to a 1-parameter family indexed by the diffusion constant $\sigma\geq 0$ \cite[App.~A.3]{Tum22}, and Bohmian mechanics is included in the family at $\sigma=0$. The theory is implausible for large $\sigma$, but $\sP_{WOD}$ will already change significantly for rather small $\sigma$.

\item Deotto and Ghirardi \cite{DG98} described further ODEs as alternatives to Bohm's. The formulas would not seem convincing as laws of nature, but they do ensure equivariance.

\item Colin, Wiseman, and Struyve \cite{CW11,Str12,MMS22} considered a $|\Psi|^2$ distributed Markov process for Dirac wave functions $\Psi$, called the ``zig-zag process'' and based on introducing a further hidden variable in $\{+1,-1\}$ for each particle representing ``handedness.'' In the non-relativistic limit for a wave function factorizing into a  spinor and a space part, $\Psi=|\vn\rangle\otimes\psi$, the particle carries out, on top of a slow Bohm-like motion, a fast (speed $c$), random, back-and-forth motion in direction $\pm\vn$ (see \cite{MMS22} for pictures).
\end{itemize}

\subsubsection{Implications}

The existence of different equivariant laws of motion shows, in fact, that \eqref{TDTWID} can fail: While the joint wave function $\Phi_\tau$ of particle and detector (and records) at a late time $\tau$ fixes $\sP_D$, different equivariant laws of motion will in general lead to different trajectories $Q(t)$ of particle and detector together, and we see no reason why in general these different trajectories should have equal $T_{WID}$ or $\sP_{WID}$.
For the zig-zag process, Nelson's stochastic mechanics, and its extensions with other diffusion constants $\sigma$, it seems rather obvious from the stochastic fluctuations that $\sP_{WID}$ will be different from what it is in Bohmian mechanics, for example allowing for many much smaller arrival times.

\section{Conclusion}
\label{sec:rem}

Das and D\"urr hoped that the experiment they studied could be realized and would yield outcome statistics in agreement with the $\sP_{WOD}$ that they computed. Our results show that Bohmian mechanics predicts otherwise. 
We have thus shown that the quantity $T_{WOD}$ is not measurable, that is, that there is no experiment whose outcome conveys the time at which the particle would have reached the surface $\Sigma$ in the absence of detectors. 
The upshot of our considerations is that arrival times are less accessible than some might imagine. 

\bigskip

\noindent{\bf Acknowledgments.} We thank Dustin Lazarovici and  Ward Struyve for useful discussions. 

\appendix

\section{The General Emergence of POVMs}
\label{app:POVM}

Consider the statement:
\begin{equation}\label{theomo}
\begin{minipage}{.7\textwidth}
{\it The statistics of the result of every quantum experiment is governed by a POVM.} 
\end{minipage}
\end{equation}
This statement encapsulates a fundamental principle in quantum mechanics. In this appendix, we will delve into the significance of this assertion, highlighting its status as a theorem in both standard quantum mechanics and Bohmian mechanics, as well as other quantum theories without observers, such as stochastic mechanics or GRW theories. Although these findings are not novel, with roots tracing back more than fifty years in standard quantum mechanics and over two decades in Bohmian mechanics, we will present a comprehensive overview covering both cases, drawing inspiration from the exposition provided in \cite{DGZ04}. As an exercise for the reader, one can also explore its applicability to other quantum theories without observers.

To elucidate the significance of \eqref{theomo},
we shall first recall the nature of a POVM as a mathematical object and elucidate why it serves as a natural generalization of a quantum observable represented by a self-adjoint operator. Next, we will expound upon the precise sense in which a POVM governs the statistics of an experiment, and finally, we will elucidate why the proof of the theorem \eqref{theomo} pertains to  {\em every} quantum experiment. This will involve highlighting the general minimal conditions that distinguish a quantum experiment from any other interaction between two quantum systems.

\paragraph{POVM} A positive-operator-valued measure (POVM) is a normalized, countably additive set  function $O$   on some value space $\Lambda$, assigning positive operators $O (\Delta)$ on a Hilbert space $\mathcal{H}$ to (measurable) subsets $\Delta$ of $\Lambda$. Being normalized means that $O(\Lambda)= I$, where $I$ is the identity operator on  $\mathcal{H}$. Standard examples for $\Lambda$ are $\mathbb{R}$, $\mathbb{R}^n$ and discrete sets.

According to the spectral theorem, self-adjoint operators are in one-to-one correspondence with projection-valued measures (PVMs) on $\RRR$, which constitute a particular class of POVMs on $\mathbb{R}$ whose positive operators are orthogonal projections on the Hilbert space $\mathcal{H}$. Moreover, when a POVM $O$ is sandwiched between normalized states $\psi\in \mathcal{H}$, it yields a probability distribution
\begin{equation}
\mu^O_\psi (\Delta) = \langle\psi , O (\Delta)\, \psi\rangle.
\label{eq:povmmv}
\end{equation}
When $O$ is a PVM and $\Delta\subset \mathbb{R}$, $O (\Delta)$ is a spectral projector associated with a self-adjoint operator $A$, and  \eqref{eq:povmmv} provides the usual probability distribution of the results when $A$ is ``measured." In this sense, POVMs are a natural generalization of the notion of a quantum observable as a self-adjoint operator. So, returning to the statement \eqref{theomo}, for the statistics of a quantum experiment to be governed by a POVM means that the probability distribution of its result $Z$ can be expressed as~\eqref{eq:povmmv}, with $\psi$ being the initial state of the system on which the experiment is performed (further elaboration on this below).

\paragraph{MVQM} The notion of POVM is mathematically equivalent to that of normalized measure-valued-quadratic-map (MVQM). A measure-valued map  on $\mathcal{H}$,  
$\psi \mapsto \mu_{\psi}$, is said to be quadratic 
if $\mu_{\psi}= B(\psi, \psi)$, representing 
 the diagonal part of a sesquilinear map $B$, from $\mathcal{H}\times\mathcal{H}$ to
the complex measures on some value space $\Lambda$. If $B(\psi, \psi)$ is a probability measure whenever $\|\psi\| =1$, 
the map is said to be normalized. The following theorem easily follows:
\begin{equation}\label{theopomv}
\begin{minipage}{.8\textwidth}
{\it Eq. \eqref{eq:povmmv} defines a
     canonical one-to-one correspondence between POVMs and normalized MVQMs on $\mathcal{H}$.}
\end{minipage}
\end{equation}
(See \cite{DGZ04} for details.)

\paragraph{General Notion of  Quantum Experiment}
Let us list the minimal conditions that are common to   {\em any} experiment across different formulations of quantum mechanics. 
\begin{enumerate}

\item A quantum experiment involves a system to be ``measured" and an apparatus. The apparatus, being macroscopic, records the information conveyed by the experiment. This information is captured as an output, which is typically represented by a macroscopic variable such as the orientation of a pointer or a stable record displayed on a computer screen.

\item  The experiment begins at an initial time, say $t=0$, at which the apparatus assumes a fairly definite initial wave function, often termed the ready state $\Phi_0=\Phi_{0}(y)$. Here,  $y$
represents a generic microscopic configuration of the apparatus, which includes the positions of particles constituting the apparatus, as well as fields or any other relevant descriptors for its microscopic characterization. Further elaboration on this point will follow.  With  $\Phi_0$ is associated a macroscopic variable representing a null orientation of the pointer.

\item The experiment is performed on a system with an initial wave function $\psi=\psi (x)$. Here, $x$ denotes the generic configuration of the system to be measured, such as the position of an electron. However, there are no restrictions on the range of possible microscopic configurations of the system. Furthermore, the wave function of the system, as well as that of the apparatus, may encompass  internal degrees of freedom, such as accounting for spin. While $\Phi_0$ is fixed as part of  the apparatus setup, $\psi$ varies,  as the apparatus is designed to probe the system for a given, but not fixed, initial system's wave function.

\item  Part of the fundamental notion of an experiment is that the system and apparatus are independent immediately before the experiment begins. Thus, the initial state of the composite formed by the system and apparatus is given by\footnote{It might be argued that assuming a sharp preparation of  $\psi $, as well as the ability to set the apparatus in a definite initial state   $\Phi_0 $, is somewhat unrealistic, as some uncertainty is unavoidable. This uncertainty, however,  can be addressed by considering quantum states to be random, following appropriate probability distributions; this is particularly true for the apparatus---a macroscopic system. This approach leads to describing quantum states in terms of suitable density matrices. Importantly, this mathematical adjustment does not change the conclusions we will reach below; see~\cite{DGZ04}.
}
\begin{equation}\label{ps}
    \Psi_0 = \psi \otimes \Phi_0 .
\end{equation}

\item As experiments inevitably conclude, let $t=T$ denote the time at which the experiment ends. If the composite system formed by the system and apparatus, with generic configuration $q=(x,y)$, is a closed system during the time interval $[0,T]$,  its state $\Psi_t=\Psi_t(q)$ evolves unitarily over this period. If $t\mapsto U_t$  is the unitary evolution operator generated by the interaction between the system and apparatus,  then the final state of the composite is
\begin{equation}
\Psi_T = U_T \Psi_0 = \Psi_T (q).
\end{equation}

\item The experiment comes equipped with a calibration,  a macroscopic function $Z= F(q)$ assigning numerical values to the outcome of the experiment. According to Born's rule, the statistical distribution of $Z$ is determined by the probability measure
\begin{equation}
\label{eq:mapsrhoz}
\rho_\psi^Z = \rho_{\Psi_T} \circ F^{-1}
\end{equation}
where  \begin{equation}
\label{eq:mahoz}\rho_{\Psi_T} =| \Psi_T (q)|^2\,. \end{equation}
\end{enumerate} 

\paragraph{The POVM Theorem for a General  Quantum Experiment}

Upon examination, it's evident that the mapping \(\psi \mapsto \rho^{Z}_{\psi}\), as given by \eqref{eq:mapsrhoz}, from the initial wave function of the system to the probability distribution of the result \(Z\), indeed constitutes an MVQM on the Hilbert space $\mathcal{H}$ of the system measured. This conclusion follows because the mapping can be viewed as a composition of maps \(\psi \mapsto \psi \otimes \Phi_0 \mapsto \Psi_T \mapsto |\Psi_T(q)|^2 \mapsto \rho_\psi^Z\), where all are linear except for \(\Psi_T \mapsto |\Psi_T(q)|^2\), which is quadratic.

Thus, by Theorem \eqref{theopomv}, there exists a POVM \(O\) such that \(\rho^{Z}_{\psi}\) can be expressed as \eqref{eq:povmmv}. (As a simple exercise, one may derive the explicit formula \eqref{POVMformula} for \(O\) that involves \(\Phi_0, U=U_T\), and \(F\), see \cite{DGZ04}.)

Therefore, the association of every experiment with a POVM, which succinctly encapsulates the statistical outcomes, is almost a mathematical triviality. This principle holds not only in standard quantum mechanics,  but also extends to Bohmian mechanics, as we will elaborate next.

\paragraph{The POVM Theorem in Bohmian Mechanics} With its increased conceptual precision, Bohmian mechanics eliminates the vague distinction between microscopic and macroscopic. While the standard formulation requires postulating that macroscopic variables such as $Z$ are always well defined, in Bohmian mechanics, they are so because they are functions of the actual microscopic variables $Q=(X,Y)$ evolving according to the guidance law.  

The initial Bohmian state \((Q_0, \Psi_0)\) evolves into the final state \((Q_T, \Psi_T)\), where the initial orientation of the pointer, determined by the initial actual configuration \(Y_0\) of the apparatus, transforms into the final outcome \(Z = F(Q_T)\) that quantifies the experimental result. Since the Born rule governs the statistics for Bohmian mechanics just as it does for standard quantum mechanics,  \(Q_T\) is distributed according to \(|\Psi_T(q)|^2\). Consequently, \(Z\) follows the distribution described by \eqref{eq:mapsrhoz}, reaffirming the POVM Theorem \eqref{theomo} within Bohmian mechanics.

The inevitability of this theorem stems from the fact that items 1--6 are necessary conditions for an experiment, both in the standard as well in the Bohmian case. In the latter case one could consider item 5 to be too restrictive. However, this is not the case, as the closedness of the $(x,y)$-composite, and thus the unitarity of its wave function evolution, can be ensured by incorporating the largest relevant composite system, and if necessary  the entire environment of the $x$-system, into $y$, which, with the $x$-system, would thus encompass the entire universe. Accordingly, $\Psi_0$ in item 4 can be regarded as the wave function of the universe at $t=0$.\footnote{We say ``can be regarded'' and not ``is'' because it would be rather unrealistic for the wave function of the universe at the initial time of the experiment to be a product state as in \eqref{ps}. However, if the difference between \eqref{ps} and the actual wave function of the universe could not be ignored, it would not be reasonable to treat $\psi$ as the wave function of the system.}

Here are additional considerations that reinforce the above conclusion:
\begin{itemize}
    \item[(a)] Since we have assumed unitarity and no additional conditions on the interaction between the system and apparatus, our analysis applies equally well within the relativistic framework, where the unitary group of time translations \( t \mapsto U_t \) is inherent in any relativistic quantum field theory,  forming part of the projective unitary representation of the Poincar\'e group. The configuration space may encompass any appropriate microscopic variables within this framework, such as local fermion numbers, as suggested by Bell \cite{Bell84} and further elaborated in \cite{DGTZ05}.

    \item[(b)] Nothing crucially depends on the specifics of the Bohmian evolution; the same holds true for other theories with entirely different trajectories, e.g., stochastic mechanics, provided the Born rule holds (and the theory allows for sufficiently stable macroscopic records).
    
    \item[(c)] Even unitarity is arguably not a necessary condition. Instead, the (approximate)  Born rule for macroscopic variables, a condition satisfied by GRW theories and spontaneous collapse models for which the evolution of the wave function is not unitary, would probably suffice \cite{AGTZ08}.
\end{itemize}

In summary, the universal applicability of the POVM Theorem \eqref{theomo} in both Bohmian mechanics and other formulations of quantum mechanics is essentially a mathematical inevitability arising from the unitarity (specifically linearity) of quantum state evolution and the Born rule. A significant departure from either of these conditions  is the only plausible way to circumvent this conclusion (to appreciate why we say ``significant," recall item (c) above).

\section{The Objection of Das and Aristarhov}
\label{app:B}

In a recent preprint \cite{DA23} Das and Aristarhov object to our use of the POVM theorem, expressing doubt that the wave function $\Psi_t$ of system and apparatus in Appendix~\ref{app:POVM} obeys ``an autonomous Schr\"odinger evolution.'' They suggest instead that its evolution, like that of a general (conditional) wave function of a system interacting with its environment, might be ``highly non-linear and  non-unitary.'' 
In this regard, we note the following:
\begin{itemize}
    \item As we indicated in Appendix~\ref{app:POVM}, $\Psi_t$ can be taken to be the wave function of the universe, which, in Bohmian mechanics, certainly evolves unitarily.
    \item More importantly, the idealization that the  system-apparatus composite (like that of any system whose behavior we want to analyze according to a particular theory) can be  sufficiently isolated from its environment to be treated as a closed system  is a standard assumption for the analysis of a system in any theory, be it Bohmian mechanics, standard quantum mechanics, or any theory whatsoever. 
    \item To design an apparatus to fulfill a particular purpose, it would seem to be necessary to treat the apparatus as a closed system. After all, if interactions with the environment remain significant and can't be ignored, the apparatus could hardly be expected to fulfill the purpose for which it was designed.
    \item Of course,  there is no guarantee that an apparatus can be so designed and so isolated as to justify the idealization required for the analysis. It might well be, for example, that the workings of the apparatus are just too sensitive to environmental disturbances.  In such a situation we would presumably have little idea of how the apparatus should behave or what it might reveal. But we could be quite confident that it would not do what it was designed to do.
    \item In this regard we note the claim of Das and Aristarhov that while the assumption for  the POVM theorem may be correct, ``it may be wrong and the actual distribution of the arrival times will resemble the one calculated in \cite{DD19}.'' They seem to be making the rather implausible 
    claim that the uncontrollable disturbances produced by the environment, which itself knows nothing directly about the system being measured,  might nonetheless undo the behavior of the idealized system-apparatus composite  and convert  results that significantly differ from the actual unmeasured arrival times to results that agree with them.
    
    \item The conspiratorial claim just described resembles that of the advocates of local hidden variables who suggest that the apparently random effects of the environment for experiments testing Bell's inequality  might produce precisely the correlations between the choices of the experimenter and the initial state of the entangled particle pair so as to produce the quantum predictions via only local mechanisms.
    
\end{itemize}

\end{document}